# Tunable Atomically Wide Electrostatic Barriers Embedded in a Graphene/WSe$_2$ Heterostructure

Hui-Ying Ren[1,2], Yue Mao[3], Ya-Ning Ren[1,2],*, Qing-Feng Sun[3,4,5], and Lin He[1,2],*

**Affiliations:**
[1]Center for Advanced Quantum Studies, Department of Physics, Beijing Normal University, Beijing, 100875, China
[2]Key Laboratory of Multiscale Spin Physics, Ministry of Education, China
[3]International Center for Quantum Materials, School of Physics, Peking University, Beijing, 100871, China
[4]Collaborative Innovation Center of Quantum Matter, Beijing 100871, China
[5]Beijing Academy of Quantum Information Sciences, West Bld. #3, No. 10 Xibeiwang East Road, Haidian District, Beijing 100193, China
[§]These authors contributed equally to this work.
[†]Correspondence and requests for materials should be addressed to Ya-Ning Ren (yning@mail.bnu.edu.cn) and Lin He (e-mail: helin@bnu.edu.cn).

**Inducing and controlling electrostatic barriers in two-dimensional (2D) quantum materials has shown extraordinary promise to enable control of charge carriers, and is key for the realization of nanoscale electronic and optoelectronic devices[1-10]. Because of their atomically thin nature, the 2D materials have a congenital advantage to construct the thinnest possible p-n junctions[1,3,7,9,10]. To realize the ultimate functional unit for future nanoscale devices, creating atomically wide electrostatic barriers embedded in 2D materials is desired and remains an extremely challenge. Here we report the creation and manipulation of atomically wide electrostatic barriers embedded in graphene/WSe$_2$ heterostructures. By using a STM tip, we demonstrate the ability to generate a one-dimensional (1D) atomically wide boundary between 1T'-WSe$_2$ domains and continuously tune positions of the boundary because of ferroelasticity of the 1T'-WSe$_2$. Our experiment indicates that the 1D boundary introduces atomically wide**

**electrostatic barriers in graphene above it. Then the 1D electrostatic barrier changes a single graphene/WSe$_2$ heterostructure quantum dot from a relativistic artificial atom to a relativistic artificial molecule.**

The p-n junctions, as the most efficient way to manipulate behaviors of charge carriers, are the essential building blocks for electronic and optoelectronic devices[1-10]. Owing to their atomic thickness, two-dimensional van der Waals (2D vdW) materials provide an unprecedented atomically thin platform for creating the thinnest possible p-n junctions[1,3,7,9,10]. Therefore, further reducing the lateral scale of the p-n junctions in 2D vdW devices is crucial to realize the ultimate functional unit for future nanoscale devices. Thus far, many different methods have been developed to create in-plane p-n junctions embedded in 2D materials and they provide spatial control down to about ten nanometers or so[11-26]. For example, recent experiments indicate that, through substrate engineering, circular graphene p-n junctions, *i.e.*, graphene quantum dots (QDs), with diameter as small as about ten nanometers can be realized in a continuous graphene sheet[13,15-17,22,24,26]. Although creating atomically wide electrostatic barriers embedded in 2D materials is desired, the experimental realization is still lacking up to now. Here, we report a route to realize atomically wide in-plane electrostatic barriers with continuously tunable sites in graphene/WSe$_2$ heterostructures. Taking advantage of ferroelasticity of the 1T'-phase WSe$_2$[27], we, for the first time, demonstrate the ability to tune positions of one-dimensional (1D) atomically sharp boundaries between 1T'-WSe$_2$ domains with nanoscale precision by using scanning tunneling microscope (STM) tip.

Our measurements indicate that the 1D boundary introduces atomically wide electrostatic barriers in graphene, which strongly alter behaviors of massless Dirac fermions in graphene. In a nanoscale graphene/WSe$_2$ heterostructure QD, our experiments supported by theoretical calculation demonstrate that the 1D electrostatic barrier changes quasibound states in the QD from relativistic artificial atomic states to relativistic artificial molecular states.

Because graphene only has one-atom-thick sheet of carbon atoms, its electronic doping depends sensitively on the supporting substrate[28], enabling the realization of high-quality p-n junctions in graphene through the substrate engineering[13,15-17,22,24,26]. In the meanwhile, the monolayer WSe$_2$ can exist in different atomic and electronic phases, including 1H, 1T, and 1T', and many methods have been developed for triggering phase transitions between these polymorphs[29-35]. Therefore, using different structural polymorphs of the WSe$_2$ as the supporting substrates can introduce different electronic doping in graphene and generate well-defined QD in graphene[22,24,26], as schematically shown in Figs. 1a, 1b, and 1d. The 1T' phase, obtained from Peierls distortion in the 1T parent phase, has three equivalent orientations, therefore, boundaries separating different 1T' domains with different orientation variants are expected to form[27,35]. According to previous study[35], the electronic properties at the boundaries are quite different from that within the 1T' domains because the existence of localized states along the boundaries. With considering the fact that the boundary is atomically sharp, it is expected to introduce atomically wide electrostatic barriers in graphene above it (Fig. 1c). By introducing the 1D atomically sharp boundary in the

graphene/WSe$_2$ heterostructure QD (as schematically shown in Fig. 1e), then it is possible to tune the electrostatic potential and, consequently, the quasibound states within a single nanoscale QD.

In our experiments, high-quality graphene/WSe$_2$ heterostructure is obtained by using transfer technology of graphene monolayer onto mechanical-exfoliated thick 2H-phase WSe$_2$ sheets[22,24,26] (see Methods section for details of fabrication). Then, a nanoscale 1T'-phase monolayer WSe$_2$ island is created at the interface of the graphene/WSe$_2$ heterostructure by using STM tip pulses. The nanoscale WSe$_2$ island is generated from the 2H-phase WSe$_2$ substrate and there is a structural phase transition during the formation of the interfacial 1T'-phase island, as reported previously[22,24,26]. Figure 2a shows a representative STM image of a phase-transitioned 1T'-phase WSe$_2$ island generated at the interface of graphene/WSe$_2$ heterostructure. The WSe$_2$ substrate is in the 2H phase and is characterized by a trigonal-prismatic coordination of transition metal atoms (see supplemental Fig. S1 for atomic-resolved characterizations). However, the interfacial WSe$_2$ island becomes monoclinic 1T' phase, in which one array of top Se atoms becomes higher than the adjacent array, as clearly shown in atomic-resolved STM measurement (Fig. 2e). Our experiment indicates that the interfacial WSe$_2$ island in the Fig. 2a is in a single 1T' domain.

A key observation in this work is that we can tune the 1T' domains with different orientation variants in the interfacial WSe$_2$ islands by using the STM tip. Figure 2 summarizes a representative result that an atomically sharp 60° domain boundary is created and moved in the interfacial WSe$_2$ island. The formation of the 60° domain boundary can be clearly identified in atomic-resolved STM images, as shown in Figs. 2f-2h. Our experiment indicates that the positions of the boundary can be continuously moved with nanoscale precision in the interfacial WSe$_2$ island (Figs. 2a-2h). The above results are further confirmed in spatial-resolved scanning tunneling spectroscopy (STS) mappings, which directly reflect local density of states (LDOS) at the measured energies. The localized electronic states of the boundary[35] result in the high intensity features along it in the STS maps, as marked by red dashed lines in Figs. 2j-2l, enabling us to image the creation and moving of the boundary in the WSe$_2$ island. In conventional STM measurements (i.e., conventional STM imaging and STS spectroscopy), the 1T' domains and their boundaries are stable in the 1T'-phase WSe$_2$ island, which allows us to characterize them systematically. However, our experiment indicates that STM tip pulses, high current raster STM scan, and movement of the interfacial WSe$_2$ island by using STM tip can lead to the formation of the boundaries between the 1T' domains and result in their movement in the interfacial WSe$_2$ island (see supplemental Figs. S2-S5 for more experimental results on different interfacial WSe$_2$ islands). According to our experiments, the tip-induced movement of the interfacial WSe$_2$ island provides the most controllable way to tune positions of the boundary, showing the ability with nanoscale precision (Fig. 2). In a previous study[27], theoretical calculations predicted the ferroelasticity of the 1T'-phase WSe$_2$ and the orientation variants of the 1T' phase were predicted to be switched by a mechanical strain. Our experimental result provides evidence of the ferroelastic nature of the 1T'-phase WSe$_2$ and implies that the tip-induced movement of the interfacial WSe$_2$ island can impose an external stress stimulus on it. Besides that, the observed tip-induced variant switching also indicates that the transformation barrier associated with the variant switching is small, as predicted theoretically in ref.[27].

Around the graphene/WSe$_2$ heterostructure QD, as shown in Fig. 2a, the supporting substrates of graphene on and off the interfacial WSe$_2$ island are the 1T'-phase and 2H-phase WSe$_2$ respectively. The different substrates will generate a large electrostatic potential on the graphene around the 1T'-phase WSe$_2$ island, which is expected to result in a sequence of temporarily-confined quasibound states in graphene[22,24,26], as confirmed in our experimental results shown in Fig. 3. Figure 3a shows a representative STS, *i.e.*, d$I$/d$V$, spectroscopic map recorded across the graphene/WSe$_2$ heterostructure QD shown in Fig. 2a. Our measurement indicates that the 1T'-phase WSe$_2$ island introduces a Coulomb-like electrostatic potential on the graphene above it (see supplemental Fig. S6) and our theoretical calculation (see Methods section in Supplemental Material for details of theoretical calculation), based on the Coulomb-like electrostatic potential, reproduces quite well the main features of our experimental result (see supplemental Fig. S7). The Coulomb-like electrostatic potential generates both atomic collapse states (ACSs) at the center of the potential field and whispering gallery modes (WGMs) locate at the edge of the potential profile[22,24,26], as shown in Fig. 3a.

To explore effects of the atomically sharp boundary on the electronic properties of the graphene/WSe$_2$ heterostructure QD, STS spectroscopic maps are measured along the solid black arrows (see Figs. 2b-2d) across the boundary embedded in the QD, as shown in Figs. 3b-3d. Strongly localized electronic states are observed along the atomically sharp boundary in the energy range < -60 meV, as reported in ref.[35]. Therefore, we can clearly identify the boundary in the interfacial island according to the STS maps, as shown in Figs. 2j-2l (see supplemental Fig. S8 for more experimental data). A notable effect of the boundary on the electronic properties of the graphene/WSe$_2$ heterostructure QD is that it introduces an atomically wide electrostatic barrier within the QD. Our STS measurements indicate that the boundary generates an electrostatic barrier with height of about 100 meV and width of about 1 nm in graphene above it (see supplemental Fig. S6). According to our experiment (Figs. 3b-3d and Fig. S8), the newly-introduced 1D atomically wide barrier does not strongly affect the quasibound states with higher angular momenta in the QD because that the height of electrostatic barrier generated by the boundary ~ 100 meV is smaller than that ~ 500 meV generated by the 1T'-phase WSe$_2$ island. Our theoretical calculation with considering the 1D atomically wide electrostatic barrier within the QD also confirms the above result (Fig. S7). However, our experiment indicates that the 1D atomically wide barrier strongly alters the behaviors of the two lowest quasibound states in the QD. This is especially obvious when the 1D barrier is moved to the center of the QD. To clearly show this, the spatial distributions of the two lowest quasibound states for two different cases are shown in Figs. 4a and 4g respectively (Fig. 4a shows the result of the single 1T' domain case, and Fig. 4g shows the result of that the 1D boundary is tuned to the center of the QD). The spatial distributions of the two lowest quasibound states seem to be changed by the 1D boundary. The differences between the two cases can be seen more explicitly in the STS maps (Figs. 4b, 4c, 4h, and 4i), which directly reflect real-space distributions of the two lowest quasibound states. Obviously, the spatial distributions of the lowest ACS and WGM in the single 1T' domain case are strongly changed by the 1D barrier. In the single 1T' domain case, the quasibound states of the QD can be described well by quantum confinement of massless Dirac fermions in the Coulomb-like potential with a large cut-off radius: the lowest ACS exhibits a maximum in the center of the QD and the WGM displays ring structure locating at the edge of the potential profile, as shown in Figs. 4a-c. Our theoretical simulations, as shown in Figs. 4d-f, based on this Coulomb-like potential can reproduce quite well the spatial distributions of the ACS and WGM observed in experiment. By introducing the 1D atomically wide potential barrier in the Coulomb-like potential, our experimental results, as shown in Figs. 4g-i, indicate that the spatial distributions of the lowest ACS and WGM become similar as bonding and antibonding molecular states of two coupled graphene QDs respectively[17,24,25]. Such a result is also reproduced well in our simulations with additionally considering the 1D atomically wide potential barrier in the Coulomb-like potential, as shown in Figs. 4j-l. The 1D atomically wide potential barrier separates the lowest quasibound states of the QD into two parts, which are coupled to form the bonding and antibonding states. Therefore, the LDOS are redistributed: the bonding state is concentrating on the center, while the antibonding state exhibits the LDOS concentrating on the two ends of the molecule, as demonstrated explicitly both in our experiment and theory. Therefore, our experiments supported by our theoretical calculation indicate that the 1D atomically wide electrostatic barrier changes the quasibound states in the QD from relativistic artificial atomic states to relativistic artificial molecular states.

In summary, we demonstrate the ability to continuously tune the position of the 1D atomically sharp boundaries between 1T'-phase $WSe_2$ domains with nanoscale precision. The 1D boundary introduces atomically wide electrostatic barriers in graphene above it, which enables us to tune the quasibound states within a single nanoscale QD. Because the ferroelasticity and the atomically sharp boundaries are universal in the 1T'-phase transition metal dichalcogenide (TMD) [27], we believe that the results obtained in this work should be broadly applicable to many graphene/TMD heterostructures. Our result gives an insight for the fabrication of atomic-wide electronic devices and the movable atomically wide electrostatic barrier is promising as an ultimate junction for manipulation of charge carriers.

**Acknowledgments**

This work was supported by the National Key R and D Program of China (Grant Nos. 2021YFA1400100 and 2021YFA1401900), National Natural Science Foundation of China (Grant Nos. 12141401, 11974050, 11921005, 12374034), and "the Fundamental Research Funds for the Central Universities" (Grant No. 310400209521). The devices were fabricated using the transfer platform from Shanghai Onway Technology Co., Ltd..

**Data availability statement**

All data supporting the findings of this study are available from the corresponding author upon request.

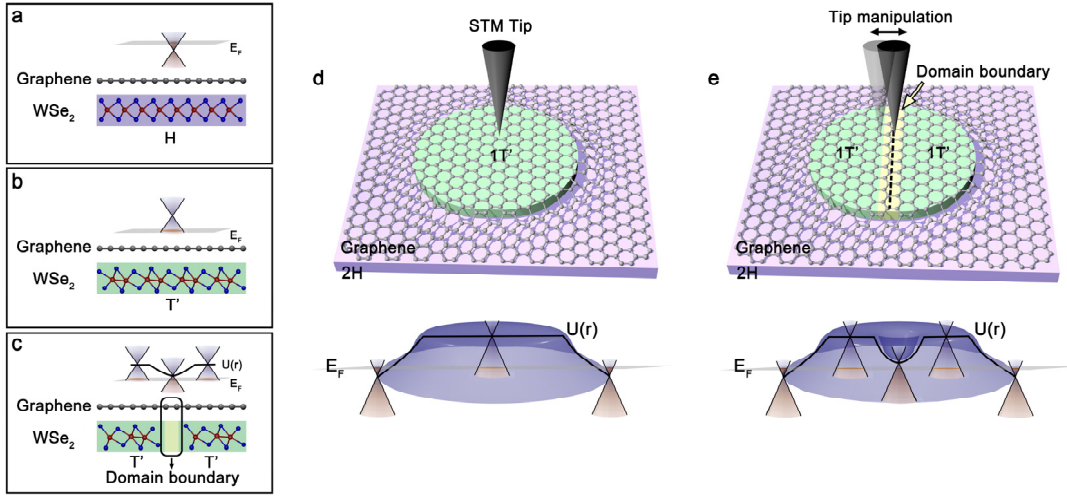

**FIG. 1. Schematics of in situ creating tunable electrostatic barriers in graphene/WSe$_2$ heterostructure. a-c.** Schematics of substrate dependence of Fermi-level shifts in graphene. The electronic doping of graphene is different on the H-phase WSe$_2$, T'-phase WSe$_2$, and boundary of the T'-phase WSe$_2$. **d and e.** Top panels: Schematics of an interfacial monolayer WSe$_2$ island in a graphene/WSe$_2$ heterostructure. An atomically sharp domain boundary with continuously tunable sites in the interfacial island can be created by using STM tip. Bottom panels: Schematics of local electrostatic potentials in graphene induced by the interfacial WSe$_2$ island. An atomically wide one-dimensional barrier is generated in graphene by the domain boundary. False colors highlight different substrates, where purple corresponding to 1H(2H)-WSe$_2$, green corresponding to 1T'-WSe$_2$, and yellow corresponding to domain boundary.

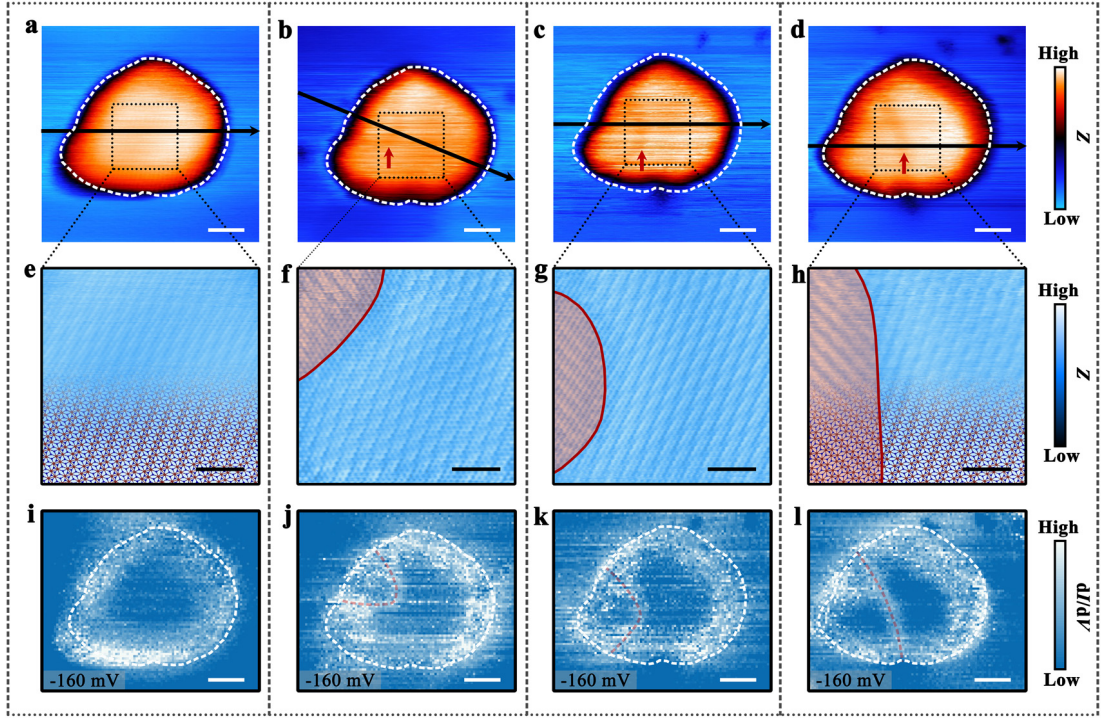

**FIG. 2. Continuously tunable domain boundary in an interfacial nanoscale 1T'-WSe$_2$ island. a-d.** STM images ($V_b$ = 400 mV, $I$ = 150 pA for panels a-c; $V_b$ = 400 mV, $I$ = 180 pA for panel d) of an interfacial 1T'-WSe$_2$ island in a graphene/WSe$_2$ heterostructure. The positions of the atomically sharp domain boundaries are marked by red arrows. **e-h.** Atomic-resolved STM images ($V_b$ = 400 mV, $I$ = 100 pA for panels e and h; $V_b$ = 400 mV, $I$ = 150 pA for panel f; $V_b$ = 800 mV, $I$ = 150 pA for panel g) in the frame of panels a-d, respectively. False colors highlight different domains with different parallel dimerization orientations, and the 60° domain boundaries are marked by red solid lines. Schematic structures of the 1T' domains are overlaid on the atomic-resolved images in panels e and h. **i-l.** Representative STS maps recorded around the interfacial 1T'-WSe$_2$ island, as shown in panels a-d, respectively, at the energy of -160 meV. The white dashed lines show outlines of the interfacial island and the domain boundaries are marked by the red dashed lines. The scale bar is 5 nm in panels a-d and i-l, and 2 nm in panels e-h.

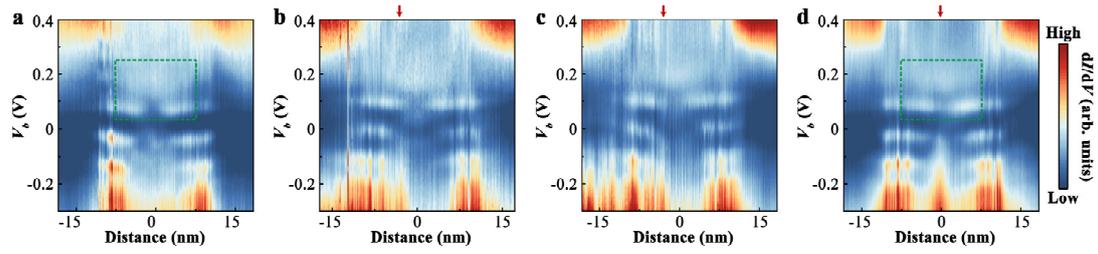

**FIG. 3. Electronic properties of the graphene/WSe$_2$ heterostructure QD.** Panels **a-d** show the d$I$/d$V$ spectroscopic maps versus the spatial position along the black solid arrows in Figs. 2a-2d, respectively. Quasibound states can be clearly observed in the QD. The red arrows mark the positions of domain boundaries. Localized electronic states are observed at the atomically sharp boundary for $V_b$ < -60 mV.

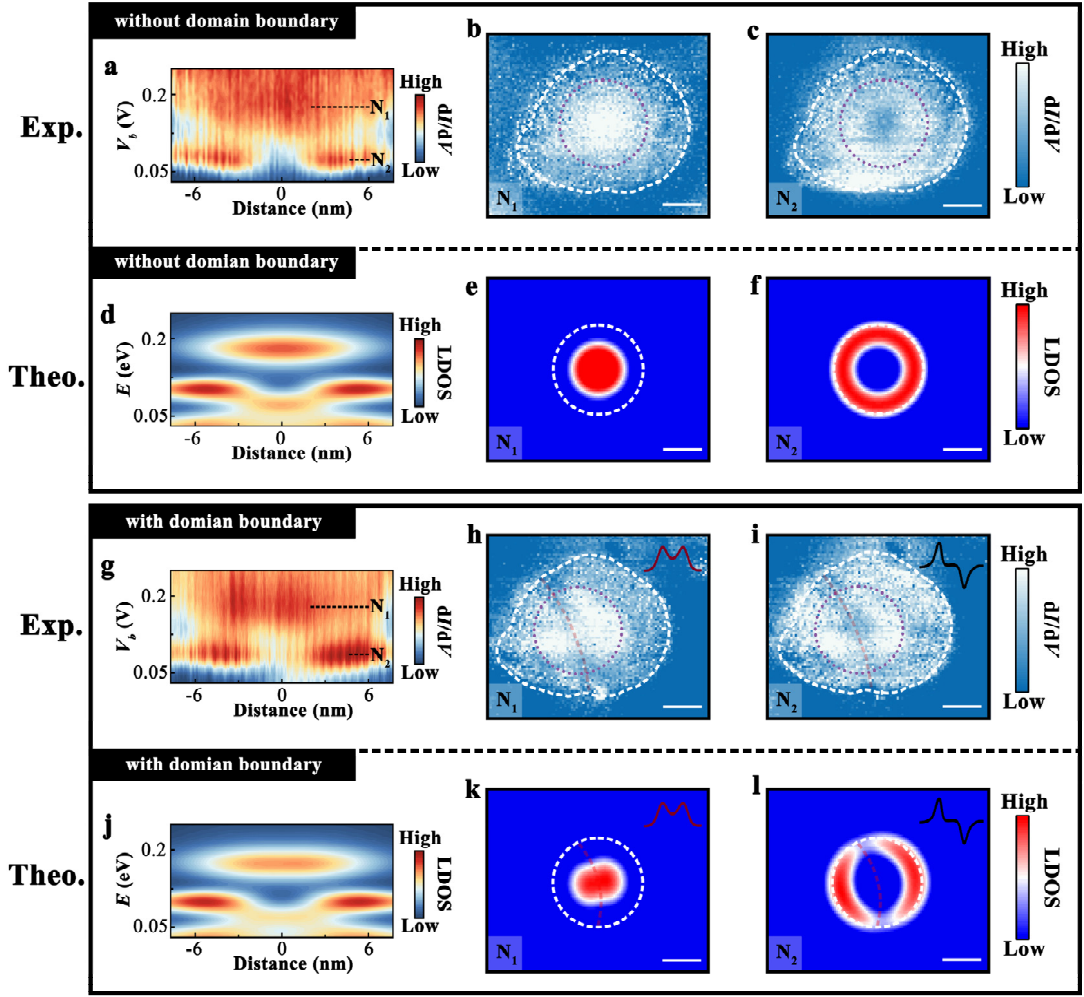

**FIG. 4. Relativistic artificial atomic states and molecular states in the graphene/WSe$_2$ heterostructure QD. a** and **g**. Zoom-in d$I$/d$V$ spectroscopic maps of green dashed rectangles in Figs. 3a and 3d, respectively. **b and c**. STS maps recorded at the two lowest quasibound states around the single domain graphene/WSe$_2$ heterostructure QD in Fig. 2a, respectively. They exhibit relativistic artificial atomic states with the ACS locating at the center of the QD and the WGM locating around the edge of the QD. **d, e, f**. The calculated LDOS space-energy maps of the QD in the single domain case. **h and i**. STS maps recorded at the two lowest quasibound states around the graphene/WSe$_2$ heterostructure QD in Fig. 2d, respectively. They exhibit relativistic artificial molecular states: the lowest ACS becomes similar as the bonding state and the lowest WGM becomes similar as the antibonding state. Insets: Schematic wave functions of the bonding and antibonding states are shown, respectively. **j, k, l**. The calculated LDOS space-energy maps of the QD with the 1D electrostatic barrier across

its center. The white dashed lines show outlines of the interfacial island and the domain boundaries are marked by the red dashed lines. The purple dashed lines in b, c, h, and i are guides to the eye. The scale bar is 5 nm.